\documentclass[aps,prl,twocolumn,showpacs,floats,floatfix,letterpaper,nofootinbib,superscriptaddress,notitlepage]{revtex4-1}
\pdfoutput=1
\usepackage{hyperref}
\usepackage{amssymb,amsmath,latexsym,mathrsfs}
\usepackage{graphicx,subfigure}
\usepackage{epsfig}
\usepackage{varioref,xr-hyper}
\usepackage{color}
\usepackage{multirow}
\usepackage{array}
\usepackage{placeins}
\hypersetup{
	colorlinks   = true,
	citecolor    = blue,
	linkcolor = red
}

\newcommand\Tstrut{\rule{0pt}{2.6ex}}

\begin{document}
	
\title{Searching for signals of inhomogeneity using multiple probes of the cosmic expansion rate $H(z)$}
\author{S. M. Koksbang}
\email{koksbang@cp3.sdu.dk}
\affiliation{$CP^3$-Origins, University of Southern Denmark, Campusvej 55, DK-5230 Odense M, Denmark}

\begin{abstract}
It is argued that cosmic chronometers yield estimates of the spatially averaged expansion rate even in a universe that is not well described by a global FLRW model - as long as the Universe is statistically homogeneous and isotropic with a sufficiently small homogeneity scale. On the other hand, measurements of the expansion rate based on observations of redshift drift will not in general yield estimates of the spatially averaged expansion rate - but it will in the case where the universe is described well by a single FLRW model on large scales. Therefore, a disagreement between measurements of the expansion rate based on cosmic chronometers versus redshift drift is an expected signal of non-negligible cosmic backreaction.
\end{abstract}	
	
\maketitle
\flushbottom
\paragraph*{Introduction}
One of the most fundamental quantities of the Universe is its expansion rate, $H(z)$. This quantity is for instance expected to be a key observable for pinpointing the precise phenomenology and hence true physical origin of dark energy. Currently, observations of the expansion rate are mainly based on cosmic chronometers and BAO measurements but in the future, several additional methods are expected to come i use.
\newline\indent
Observations are mainly interpreted in terms of the $\Lambda$CDM model and modest extensions of it - more generally, the Friedmann-Lemaitre-Robertson-Walker (FLRW) spacetime which is a solution to Einstein's field equations under the assumption of an exactly spatially homogeneous universe. However, it is well-known that the large-scale (``average'') evolution of an inhomogeneous universe in general deviates from FLRW evolution \cite{fluid1,bc_fluidII}. This effect is known as cosmic backreaction. It has been noted that backreaction can mimic dark energy and hence provide a physical explanation for the ``$\Lambda$'' in the $\Lambda$CDM model (the ``backreaction conjecture'', \cite{conjecture1,conjecture2,conjecture3,conjecture4}). A less radical possibility is that backreaction plays a smaller role in our universe, but nonetheless cannot necessarily be neglected when interpreting the precise and ample data from upcoming surveys. Which of these possibilities is correct is still up for debate; a realistic quantification of backreaction is still lacking since such a quantification requires taking non-linear relativistic effects realistically into account. Meanwhile, establishing the impact of backreaction on observations is necessary for obtaining trustworthy quantification of cosmological parameters. This includes when e.g. constraining the dark energy equation of state but it is also highly relevant when attempting to resolve e.g. the $H_0$-tension by modifying the $\Lambda$CDM model. Indeed, tensions in data are driving standard cosmology towards a potential crisis \cite{crisis1,crisis2,crisis3, crisis4} which it has even been argued may be directly related to backreaction \cite{bc_h0_tension}.
\newline\newline
Identifying observational signatures of backreaction can provide clues regarding the extent to which backreaction should be expected to affect observations. In this letter it is discussed how direct measurements of the expansion rate can be utilized to identify a signal of backreaction.
\newline

\paragraph*{Dynamics of an inhomogeneous universe}
For an inhomogeneous universe containing dust and a cosmological constant, the spatially averaged expansion rate of the flow-orthogonal foliation of spacetime fulfills the relation \cite{fluid1}
\begin{align}\label{eq:H}
	\frac{H_D^2}{H_{D,0}^2} = \Omega^D_{m,0}a_D^{-3} + \Omega_{\Lambda,0} -\frac{Q_D}{6H_{D,0}^2}  -\frac{R_D}{6H_{D,0}^2},
\end{align}
where a subscript zero implies evaluation at present time and $D$ indicates averaging over a spatial domain $D$ large enough to cover the homogeneity scale. Scalar averages are defined as $x_D:=\int_D x\sqrt{|\rm g|}/\int_D\sqrt{|\rm g|} $, where g is the determinant of the spatial part of the metric tensor. $H_D := \dot a_D/a_D$ is the spatial average of the local fluid expansion rate and $a_D:=\left(V_D/V_{D,0} \right) ^{1/3}$ is the volume scale factor normalized to 1 at present time.
\newline\indent
Comparing with the Friedmann equation, this equation has one extra term, namely the (kinematical) backreaction, $Q_D$, describing the difference between the averaged shear and the variance in the local expansion rate. In addition, the curvature term $R_{D}$ may evolve differently than $\propto a_D^{-2}$ (which is how the FLRW curvature evolves). These differences from FLRW dynamics mean that the average evolution of an inhomogeneous universe may deviate from that of an FLRW universe. To what extent this effect occurs in our universe is still undetermined but arguably, it can be expected to be of percent order and hence affect future upcoming observations at a detectable level \cite{adamek}.
\newline\newline
It was assessed in \cite{light1,light2} that the redshift along a light ray in a statistically homogeneous and isotropic universe with the typical time scale of structure evolution long compared to the travel time of light is given by
\begin{align}\label{eq:zD}
	z\approx z_D := \frac{1}{a_D}-1
\end{align}
up to statistical fluctuations. This is in agreement with studies of light propagation in concrete inhomogeneous cosmological models \cite{2region,Hellaby}\footnote{There are also studies that could be mistaken as contradicting the result, but these violate the basic assumptions underlying the result or are based on models with clear pathologies affecting the result: In \cite{tardis}, the studied model contains pathologies in the form of surface layers which significantly affect light propagation. In \cite{DR} opaque regions are introduced and this scenario is not addressed by the analyses of \cite{light1,light2}. In addition, the model has delta-function contributions to the expansion rate. In \cite{towards} the studied model describes inhomogeneities which evolve quickly compared to the time it takes a light ray to traverse the homogeneity scale.}. Thus, it is expected that the relation between the volume scale factor and the (mean) redshift is analogous to the FLRW relation.
\newline\indent
The analyses of \cite{light1,light2} also indicate that the mean redshift-distance relation can be approximated through spatially averaged quantities according to
\begin{align}\label{eq:DA}
	H_D\frac{d}{dz_D}\left[ \left(1+z_D \right)^2H_D\frac{dD_A}{dz_D} \right] =-4\pi G_N \rho^D_m D_A,
\end{align}
again in agreement with concrete examples \cite{2region,Hellaby}.
\newline

\begin{figure*}
	\centering
	\subfigure[]{
		\includegraphics[scale = 0.5]{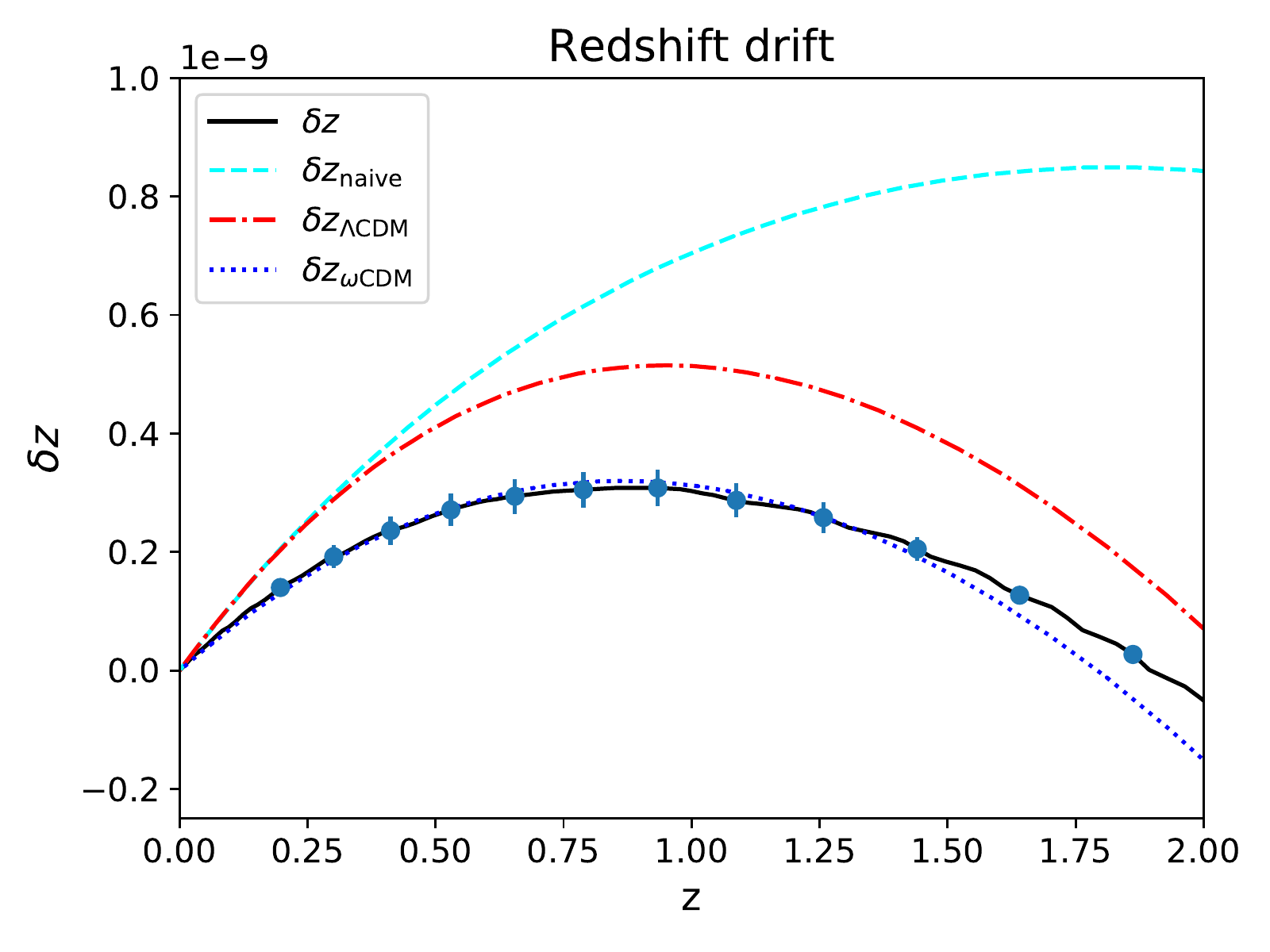}
	}
	\subfigure[]{
		\includegraphics[scale = 0.5]{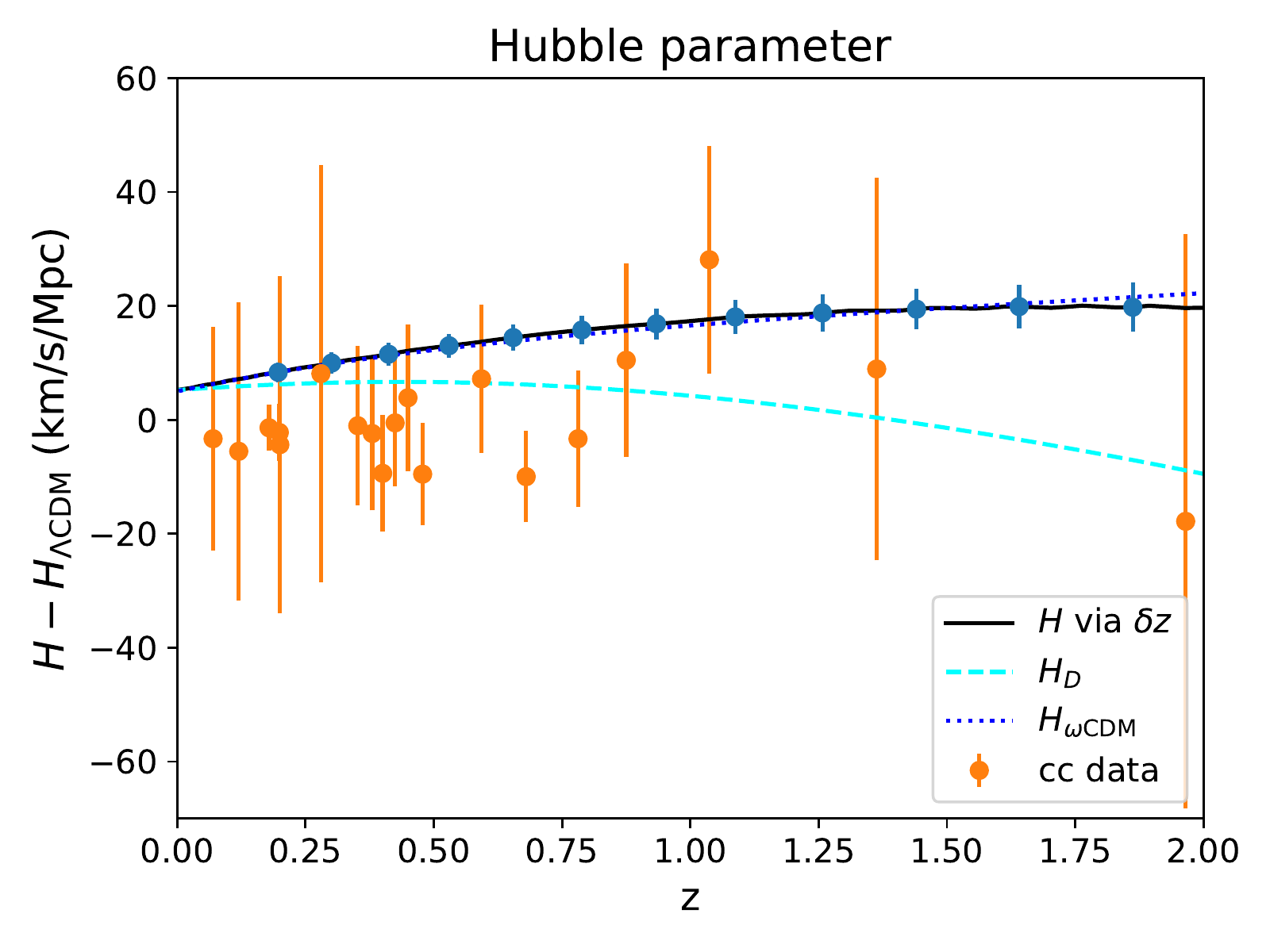}
	}
	\caption{Redshift drift and expansion rate of inhomogeneous model, $\omega$CDM model and a $\Lambda$CDM model with $\Omega_{\Lambda,0} = 0.7, \Omega_{m,0} = 0.3$ and $H_0 = 70$km/s/Mpc. $\delta z$ was computed with $\delta t_0 = 30$ years. The results based on the direct computation of $\delta z$ (no subscript in legend) along a light ray in the inhomogeneous model is shown with $10\%$ error bars. The corresponding error bars on ``$H$ via $\delta z$'' are tiny, indicating that the primary source of error on this quantity comes from errors in $z$ and $H_0$ which are at percent level. In the figure, error bars of $2\%$ of $H$ have been included to illustrate the smallness of the error bars one can expect on this type of $H(z)$ measurement. $\delta z_{\rm naive}$ is based on eq. \ref{eq:naive}. Cc data from \cite{cc1,cc2,cc3,cc4,cc5} is also shown (based on the BC03 stellar evolution model). Note that $H(z)$ obtained from $\delta z$ and for the $\omega$CDM model are largely indistinguishable.}
	\label{fig:dz_and_H}
\end{figure*}

\begin{figure}
	\centering
	\includegraphics[scale = 0.5]{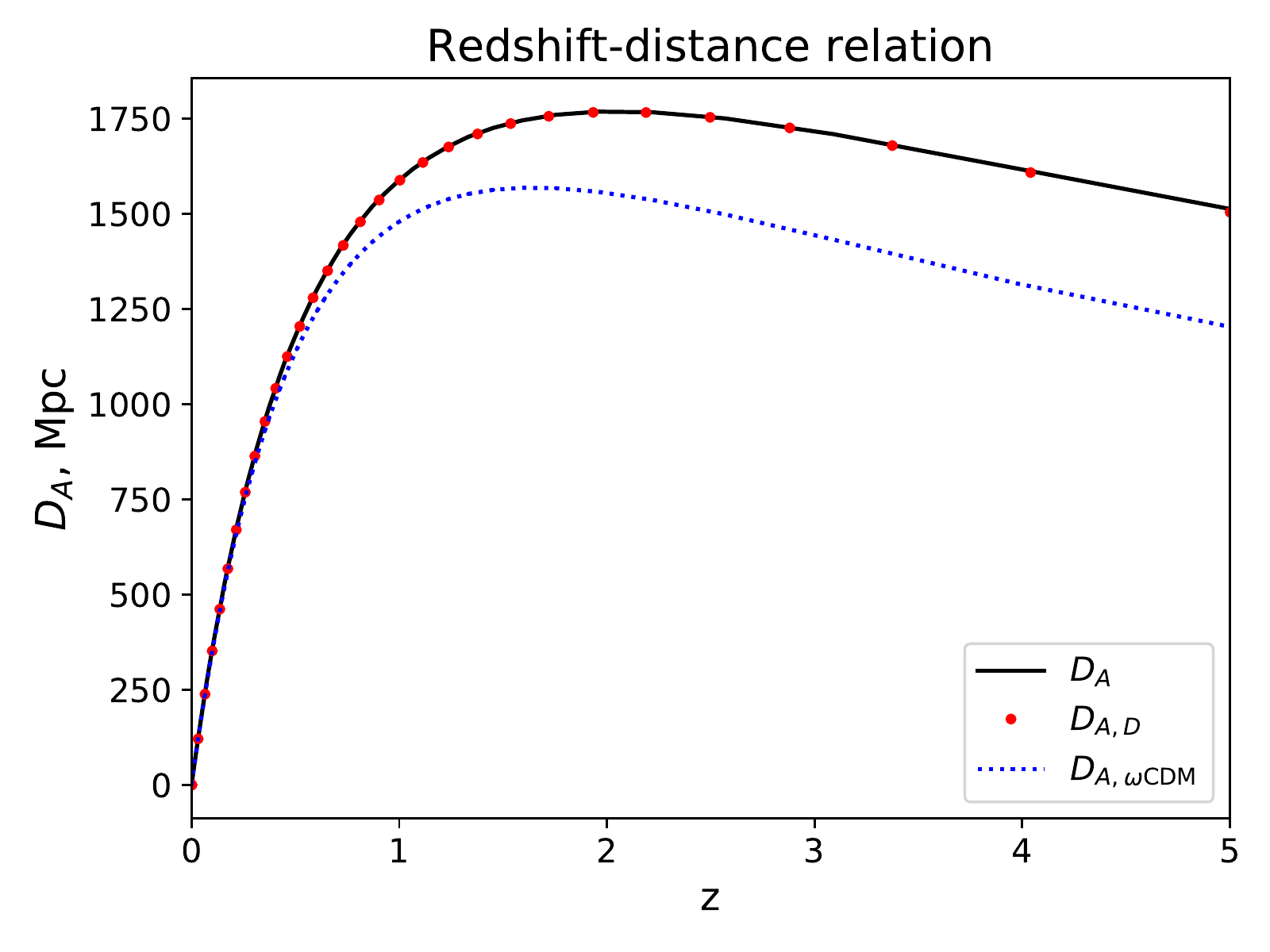}
	\caption{Redshift-distance relation according to the $\omega$CDM model and of the inhomogeneous model. For the inhomogeneous model, the relation is computed both directly along a light ray as well as based on spatially averaged quantities - the two methods yield the same result.}
	\label{fig:DA}
\end{figure}

\paragraph*{Direct measurements of the expansion rate}
As mentioned, direct measurements of the expansion rate is a promising probe for e.g. constraining the equation of state of dark energy and several different methods for ``observing'' $H(z)$ have been proposed. In an FLRW universe, these methods all measure the same quantity, but as argued below that is not the case in an inhomogeneous universe that cannot on large scales be described by a single FLRW model.
\newline\newline
{\bf Cosmic Chronometers:} Currently, most direct measurements of $H(z)$ have been obtained using cosmic chronometers (cc) which are passively evolving galaxies whose relative ages can be determined \cite{cc0} (see e.g. \cite{cc_theory1,cc_theory2,cc_theory3} for background theory and the data references in the caption of figure \ref{fig:dz_and_H} for examples). With this method, groups of cc each within a narrow redshift interval typically of order $0.1$ are identified and an ``effective'' redshift is attributed each group such that redshift differences, $\Delta z$, between the groups can be estimated. The effective relative age of the galaxies in each group is then determined, yielding the age difference, $\Delta t$, between each group. $H(z)$ can then be computed according to
\begin{align}
	\frac{\Delta z}{\Delta t}\approx \frac{dz}{dt} = \frac{d}{dt}\left( \frac{1}{a}-1\right) =-(1+z)H.
\end{align}
In an inhomogeneous universe, the redshift of each galaxy is equal to $z_D$ up to statistical fluctuations. When computing the effective redshift of each galaxy group the fluctuations will presumably be washed out since fairly large redshift intervals are used. Regardless, the observation is based on constructing a part of the $(z,t)$ curve and differentiating it. Since the total curve $(z,t)$ is well described by $(z_D,t)$, the cc measurements directly yield a measure of the spatially averaged expansion rate; according to eq. \ref{eq:zD}, we have
\begin{align}
	\frac{\Delta z}{\Delta t}\approx \frac{dz_D}{dt} = -(1+z_D)H_D.
\end{align}
Most important for obtaining this result is that the differential of the redshift is not obtained by considering the time-change in the redshift of a single object, but by estimating a function $z(t)$ by comparing $(z,t)$ at different points along an imagined $z,t$-curve.
\newline\newline
{\bf Redshift drift:} A promising (future) method for measuring $H(z)$ is through the time-change of the redshift of an object due to cosmic expansion. In an FLRW model, this so-called redshift drift, $\delta z$,  is given by \cite{sandage,mcVittie}
\begin{align}\label{eq:dz_FLRW}
	\frac{\delta z}{\delta t_0} \approx \frac{dz}{dt_0}= (1+z)H_0 - H(t),
\end{align}
where $\delta t_0$ is the observation time typically expected to be 10-100 years (usually 10-30 years are considered).
\newline\indent
Redshift drift is expected to be measured in the redshift range $2<z<5$ with CODEX \cite{CODEX1,CODEX2,CODEX3,CODEX4}. It is also expected to be obtained in the range $0<z<2$ to percent precision with SKA2 while SKA1 measurements will only with an unrealistically long observation time reach a $10\%$ precision \cite{SKA}. Note that SKA2 observations of $\delta z$ lie in the same range as cc observations.
\newline\newline
Eq. \ref{eq:dz_FLRW} is in good agreement with what one finds even if inhomogeneities are presented through perturbation theory \cite{dz_pert} and in some Swiss cheese models (see e.g. the appendix of \cite{average_zdrift}). Naively then, one would expect that the correct equivalent expression in an inhomogeneous universe would be \cite{average_zdrift}
\begin{align}\label{eq:naive}
	\frac{\delta z}{\delta t_0}\approx \frac{\delta z_D}{\delta t_0} = (1+z_D)H_{D,0} - H_D.
\end{align}
However, as shown in \cite{2region,Hellaby,Asta_dz,Asta_dz2} this expression is not correct in general for an inhomogeneous universe even if it has statistical homogeneity and isotropy on a reasonable scale. In other words, in general, the mean redshift drift is not equal to the drift of the mean redshift. In a universe that cannot globally be described by a unique FLRW model, one therefore has $ \left\langle  dz/dt_0\right\rangle\neq  dz_D/dt_0$, where triangular brackets are used to imply the mean of the quantity after averaging over several observations (to smooth away statistical fluctuations). This implies that the expansion rate obtained from $\delta z$ measurements using eq. \ref{eq:naive} (or, equivalently, eq. \ref{eq:dz_FLRW}) will not, in this type of universe, yield the spatially averaged expansion rate. Thus, {\em a signal of non-negligible backreaction is given by a mismatch between $H(z)$ obtained through cc and $\delta z$ measurements}. Before elaborating this point, comments will be given regarding other methods for measuring $H(z)$.
\newline\newline
{\bf Gravitational waves:} Perhaps the most well-known cosmological use of gravitational waves is as standard sirens to determine their luminosity distance. In connection with this, gravitational waves can be used to measure $H(z)$ directly. A method for this was proposed in \cite{GW1} based on an idea originally aimed at distant supernovae data \cite{GW2}. The method is based on an expansion of the luminosity distance leading to 
\begin{align}
	d_L^{(1)} = \frac{|v_0|(1+z)^2}{H(z)},
\end{align}
where the left hand side is the dipole of the luminosity distance and $v_0$ is the peculiar velocity of the observer. This expression is derived by assuming that the luminosity distance at lowest (monopole) order is given by
\begin{align}
d_L^{(0)}(z) = (1+z)\int_0^z \frac{dz'}{H(z')}.
\end{align}
Besides not being correct in a non-flat FLRW model, this expression is not correct in a universe containing non-negligible backreaction (as seen by eq. \ref{eq:DA} - see e.g. also the end of section 3 in \cite{light1}). A clear understanding of what a measurement of $d_L^{(1)}$ actually yields in an inhomogeneous universe requires a careful investigation beyond the scope of the current letter (but note that such an investigation might build upon the considerations presented in \cite{Asta_dipole}). However, there is no reason to expect that the relation between a monopole and dipole term of the luminosity distance will lead to the above expression if there is significant backreaction. Hence it is {\em a priori} not expected that this type of measurement will directly yield the spatially averaged expansion rate in an inhomogeneous universe. 
\newline\newline
Gravitational waves can also be used to measure $H(z)$ through a redshift drift type of observation  \cite{GW3,GW4}. In this case, the extracted $H(z)$ will not correspond to $H_D(z)$ in an inhomogeneous universe without a unique FLRW background.
\newline\newline
{\bf BAO:} Lastly, another method currently in use for measuring $H(z)$ is based on BAO (Baryon acoustic oscillations). Determinations obtained from here have some model dependence and it is not entirely clear how results from these analyses can be interpreted in an inhomogeneous universe. It will not be discussed further here, but see e.g. \cite{Asta_BAO} for some considerations regarding the BAO in an inhomogeneous universe.
\newline
\paragraph*{Numerical example}
\begin{table}
	\centering
	\begin{tabular}{c c c c c c}
		\hline\hline\Tstrut
		Model & $t_0$ (Gyr) & $H_0$ (km/s/Mpc) & $\Omega_{m,0}$ & $\Omega_{de,0}$ & $\omega_{de}$\\
		\hline\Tstrut
		Region 1& 70.5 & 5 & 10 & 0.7 & -1\\
		Region 2 & 20.2 & 70 & 0 & 0.7 & -1 \\
		Average & 13.46 & 75.2 & 0.157 & 0.437 & -1\\
		$\omega$CDM & 12.15 & 75 & 0.285& 0.67 & -0.8\\
		\hline
	\end{tabular}
	\caption{Model specifications. The average model is obtained by combining regions 1 and 2 with the ratio $0.5$ at $t = t_0$. Note that present time, $t = t_0$, is set by hand in the average region with quantities normalized accordingly.}
	\label{table:models}
\end{table}
In the event that backreaction is fairly small but non-negligible its effect on e.g. $\delta z$ may be confused with signals of dark energy not being a cosmological constant. However, for a universe based on the FLRW solutions, the corresponding $H(z)$ will be in agreement with that obtained through cc - unlike if there is significant backreaction. A comparison of $H(z)$ extracted from $\delta z$ and cc data can therefore be used to identify if backreaction has an (even small) effect on observations - an effect which could otherwise be mistaken as a signal that e.g. the dark energy equation of state parameter deviates from -1.
\newline\indent
This can be illustrated with a toy-model consisting of two different FLRW models glued together (non-smoothly) to form an inhomogeneous cosmological model where light rays are propagated alternately through each model/region. If the alternation is fitted to a certain volume fraction between the two types of FLRW regions, spatial averages of e.g. the expansion rate of the inhomogeneous model can be computed and the resulting redshift and distance measures will correspond to mean observations made in such a universe up to minor local fluctuations (see \cite{2region} for details). Here, such a model is considered with both regions containing a cosmological constant but with one region being otherwise empty and the other containing also dust. Model specifications are given in table \ref{table:models}.
\newline\newline
Figure \ref{fig:dz_and_H} shows $\delta z$ computed along a light ray in the inhomogeneous model together with $\delta z$ according to the naive expression of eq. \ref{eq:naive}, and of a $\Lambda$CDM and $\omega$CDM model (an FLRW model containing dust, curvature and a dark energy component with constant equation of state parameter $\omega_{de}\neq -1$). The $\omega$CDM model has been specifically fine tuned to have $\delta z$ and $H_0$ close to that of the inhomogeneous model (model details are given in table \ref{table:models}). As seen, if $\delta z$ is measured to $\sim 10 \%$ precision these two models will be indistinguishable for the main part of the redshift interval probed by SKA2.
\newline\indent
Figure \ref{fig:dz_and_H} also shows a comparison of different expansion rates. There is a ``direct'' expansion rate which was computed by wrongfully applying eq. \ref{eq:naive} to the direct measurements of $\delta z$. This expansion rate is what one would extract from $\delta z$ measurements and will of course agree very well with that of the $\omega$CDM model. However, the actual average expansion rate of the inhomogeneous model is also shown and it clearly differs from the former two expansion rates. This is the quantity which would be extracted from cc data. Thus, such a comparison of expansion rates obtained using $\delta z$ and cc would clearly make it possible to exclude the $\omega$CDM model in favor of the inhomogeneous model (or vice versa).
\newline\indent
Figure \ref{fig:dz_and_H} includes $H(z)$ measurements from current cc data with error bars. These error bars are quite large, but as cosmic chronometers is a fairly new probe, it is expected that the error bars (and uncertainties in systematics) will be significantly reduced by the time $\delta z$ measurements are obtained. Error bars are also included on the direct measurement of $\delta z$ in the inhomogeneous model. This error was set to $10\%$ even though SKA2 measurements (optimistically) will reach $1\%$. This was chosen simply because the error bars are so small they are barely visible even when they are chosen as large as $10\%$. The error bars in the corresponding $H(z)$ measurement was set to $2\%$ as explained in the figure text.
\newline\newline
The studied inhomogeneous and $\omega$CDM models deviate in other ways than as described above. This makes them distinguishable with different types of observations. For instance, their redshift-distance relations are quite different as illustrated in figure \ref{fig:DA}. However, the models were fine tuned specifically to have similar $H_0$ and $\delta z$ in the interval $0<z<2$ only, solely to emphasize the main point of this letter: That even a small amount of backreaction which in terms of e.g. $\delta z$ measurements can be mistaken for e.g. a signal of $\omega_{de}\neq -1$, will exhibit a clear signal by leading $H(z)$ values obtained from $\delta z$ and cc to disagree. It is easily conceivable that some of the several more complicated dark energy models studied in the literature could be fine tuned to be similar to a more sophisticated inhomogeneous model in terms of a more varied set of observations than the two focused on here. In such a case though, the homogeneous (FLRW) model can still be distinguished from the inhomogeneous one by comparing the expansion rates obtained with cc and $\delta z$ measurements.
\newline

\paragraph*{Discussion and conclusions}
In this letter it has been argued that the expansion rates, $H(z)$, obtained through different observational methods actually describe different quantities if there is non-negligible cosmic backreaction. It is unclear what several of the methods for ``directly'' observing $H(z)$ really measure in such a universe but it is argued that cosmic chronometer data will yield an estimate of the spatially averaged expansion rate while redshift drift measurements will emphatically not measure this.
\newline\indent
Since a discordance between $H(z)$ measured through cosmic chronometers and redshift drift should not occur in an FLRW universe, the comparison of these two constitute an FLRW consistency test. Other such tests have been proposed, e.g. in \cite{sum_rule,parallax, copernican}. These tests are based on deliberate data combinations made explicitly for the purpose of testing the consistency with FLRW predictions. On the other hand, with or without awareness of its possible relation to backreaction, $H(z)$ {\em will} be measured with the different methods discussed here, and the results {\em will} be combined. A general awareness of the connection to backreaction during data interpretation will increase the credibility of results extracted from such combined data sets and it will lead to a more thorough understanding of what clues the data sets hold regarding quantifying backreaction in our universe.
\newline\indent
Lastly, one may note that while a deviation between $H(z)$ measurements from redshift drift and cosmic chronometers signals non-negligible backreaction, the method as discussed here does not provide direct constraints on concrete formulas for $Q_D$ or $R_D$. It is possible that this could be remedied e.g. through further considerations along the lines of \cite{Asta_dz,Asta_dz2}. However, a general obstacle for observationally constraining $Q_D$ and $R_D$ beyond order of magnitude estimates is a lack of theoretical prediction for the evolution of these quantities. It may be possible to remedy this by e.g. using machine learning techniques combined with theoretical models to learn about likely parameterization of $Q_D$ and $R_D$. With such parameterizations at hand, the method discussed here as well as other FLRW consistency relations could be used together with e.g. Bayesian model comparison to select for inhomogeneous cosmologies.
\newline\newline
\paragraph*{Acknowledgments}
The author thanks Asta Heinesen for communications as well as the referees for their useful comments (including, but not limited to, a suggestion to change the title to nearly its current form). The author is supported by the Carlsberg foundation.

\end{document}